\documentclass[a4paper,11pt]{article}

\usepackage{jheppub} 

\usepackage[T1]{fontenc} 

\newcommand{\ep}{\epsilon}

\newcommand{\be}{\begin{equation}}
\newcommand{\ee}{\end{equation}}
\newcommand{\bea}{\begin{eqnarray}}
\newcommand{\eea}{\end{eqnarray}}

\definecolor{Red}{rgb}{1.,0.,0.}

\title{\boldmath On the light quark mass effects in Higgs boson production in gluon fusion 
}

\author[a]{Kirill Melnikov}
\author[b,a]{and Alexander Penin}

\preprint{ALBERTA-THY-01-16, TTP16-007}

\affiliation[a]{Institut f\"ur Theoretische Teilchenphysik,
Karlsruhe Institute of Technology, 76128 Karlsruhe, Germany
}
\affiliation[b]{Department of Physics, University of Alberta, 
Edmonton, Alberta T6G 2J1, Canada
}

\emailAdd{kirill.melnikov@kit.edu}
\emailAdd{penin@ualberta.ca}

\abstract{
Production of  Higgs bosons  at the LHC is
affected by the contribution of light quarks,  that mediate  the  $gg \to
Hg$ transition.   Although  their  impact is suppressed  by
small Yukawa couplings, it is enhanced by large logarithms of the ratio of the 
 Higgs boson mass or  its transverse momentum  to   light quark masses. We study
the origin  of this enhancement, focusing  on  the   abelian corrections to $gg \to Hg$ 
amplitudes of the
form $(C_F\alpha_s {\cal L}^{2})^n$, where ${\cal L} \in \{\ln(s/m_b^2), \ln(
p_\perp^2/m_b^2)\}$.  We show how these non-Sudakov double logarithmic  terms
can be resummed to all orders in the strong coupling constant. Interestingly, we find that the
transverse momentum dependence of these corrections is very weak due to a peculiar
cancellation between different logarithmic terms. Although the abelian part of QCD corrections is 
not expected to be dominant,  it can  be used to 
estimate missing higher-order corrections to light quark contributions 
to Higgs boson production at the LHC. 
}

\begin{document}
\maketitle
\flushbottom

\section{Introduction}

The  Higgs boson discovered at the LHC by ATLAS and CMS collaborations almost
four years ago \cite{aad2012gk,chat:2012gu} is a mysterious particle. Indeed, it
seems to  fit perfectly into the Standard Model (SM) of particle physics and its
mass is numerically close to the weak scale $v$. However, the mechanism that would
tie these two quantities together in a more general theory requires presence of
other, relatively light, particles that couple to the Higgs boson. Such
particles  have not been observed so far and limits on their masses gradually become so
tight that the ``natural'' relation  $m_H \sim v$  is endangered. Further
exploration of Higgs boson properties, including its couplings and quantum
numbers, will be essential for understanding to what extent the observed
particle is indeed described by the Standard Model  and, hopefully, for
discovering clues as to what the mass scale of physics beyond the Standard Model
 can actually be.

An important observable in Higgs physics is the Higgs boson transverse momentum
distribution. There are several reasons for that. On one hand, precise knowledge
of  the Higgs boson $p_\perp$-distribution is important for understanding
jet-vetoed cross sections and, more generally, observables subject to experimental  
constraints.  The uncertainties in modeling the $p_\perp$-distribution affect 
values of the Higgs coupling constants extracted from such fiducial  quantities. Since  the total
inclusive Higgs boson production cross section is currently known through  
next-to-next-to-next-to-leading order in perturbative QCD
\cite{Anastasiou:2015ema}, the uncertainty in the Higgs  $p_\perp$-distribution may 
become  the dominant one  when  future  experimental data 
is confronted with theoretical
predictions for the Higgs boson production.

Further  motivation for the precise description of the Higgs boson $p_\perp$-spectrum
comes from the observation that   the  $p_\perp$-distribution is, potentially,  a good
observable for detecting relatively light ($m \sim m_H$) colored  particles 
that couple to the Higgs boson~\cite{Arnesen:2008fb}. Indeed the  contribution of  a particle with the
mass $m \sim m_H$  to the Higgs boson production in gluon fusion  is almost independent of
$p_\perp$ for $p_\perp < m$ while for $p_\perp > m$  it rapidly decreases.
Thus the $p_\perp$-distribution of Higgs
bosons  or jets recoiling against it, may serve as a sensitive probe of this type
of physics beyond the Standard Model.

High-precision  theoretical predictions  for Higgs boson
$p_\perp$-distribution within the Standard Model  are necesary  to pursue this
program \cite{Azatov:2013xha,Langenegger:2015lra}.  Unfortunately, despite
significant progress in understanding the Higgs $p_\perp$-spectrum in recent
years, the overall situation is unsatisfactory. The main challenge 
is to describe  the bottom quark contribution to  the Higgs boson production in gluon fusion 
at moderate values of the transverse momentum.
  Indeed, the $gg \to H$ transition in the Standard Model is dominated 
by the top-quark loop, thanks to the large Higgs-top Yukawa coupling. 
 Since the top quark mass is large compared to the
Higgs mass,   it is possible to integrate out the top quark  and  describe the Higgs
production at sufficiently low transverse momentum in the effective field theory
with a local  $ggH$ interaction.  This  reduces the number of loops
in perturbative  computations by one and allows us to push them to very high 
orders in QCD perturbation theory.  Within this approximation, the Higgs
$p_\perp$-distribution has been evaluated  through next-to-next-to-leading order
at high $p_\perp < m_t $ \cite{Boughezal:2015dra,Boughezal:2015aha} and to 
next-to-next-to-leading logarithmic accuracy at low $p_\perp < m_H$ \cite{deFlorian:2011xf,Becher:2012yn}.\footnote{For 
a recent discussion and further references see Ref.~\cite{Neill:2015roa}.}

At the same time understanding  the bottom-quark
contribution to $gg \to Hg$ turned out to be  more involved.\footnote{Contributions of even lighter quarks are 
negligible.}  Indeed, since $m_b \sim 4.2~{\rm GeV}$, the
point-like approximation for the bottom quark contribution to $ggH$ vertex is only
valid for tiny transverse momenta $p_\perp < m_b$. In a broader
and more interesting momentum region  $p_\perp > m_b$, the local vertex
approximation for the bottom quark-mediated  $ggH$ interaction is invalid and we must  deal with the
computation of complicated   box diagrams with internal masses. Calculation  of
such diagrams at two and more  loops is beyond the reach of existing  computational
techniques. As the result, the $gg \to Hg$ amplitudes for $p_\perp > m_b$ are
only known in the  leading (one-loop) approximation.

The bottom quark contribution to Higgs boson production is small, compared to the contribution of the top quark.
However, it is still relevant  phenomenologically
because of  the high  precision of forthcoming  experimental measurements
of  the Higgs-gluon coupling  {\it and} because  the bottom quark  contribution is
dynamically enhanced. Indeed, although the coupling of the bottom quark to the Higgs
boson is small compared to the Higgs-top coupling,     
the $n$-loop bottom quark contribution to $gg \to Hg$ is enhanced by two powers 
of large logarithms per one power of $\alpha_s$,  {\it i.e.} 
${\cal O}(\alpha_s^n {\cal L}^{2n})$, where ${\cal L} \in \{
\ln(m_H^2/m_b^2),\ln(p_\perp^2/m_b^2) \}$.  For relevant  values of  the transverse
momentum $p_\perp \sim 30~{\rm GeV}$ and the Higgs boson mass $m_H = 125~{\rm GeV}$,
these logarithms can be numerically quite large  ${\cal L}^2 \sim 20 - 50$.  In
fact, the magnitude of the double logarithmic corrections suggests that the
all-order resummation may be necessary.

The origin of these logarithmically enhanced terms  is currently not well
understood. Although their double logarithmic nature suggests a mechanism 
similar to the Sudakov enhancement \cite{Sudakov:1954sw},  as we explain below 
the mass suppression of the amplitude 
${\cal M}_{gg \to Hg} \sim m_b^2$ makes such an interpretation problematic.
Contribution of bottom quarks  to the Higgs boson production in gluon
fusion was discussed in
Refs.~\cite{Mantler:2012bj,Grazzini:2013mca,Banfi:2013eda} in the context of
$p_\perp$-resummation.  There it was pointed out that the standard
technology  of $p_\perp$-resummation  only applies  for  $p_\perp < m_b$,
while  for larger values of $p_\perp$ it is  incomplete. 
The authors of Refs.~\cite{Mantler:2012bj,Grazzini:2013mca,Banfi:2013eda} then used
differences  between various resummation prescriptions  to  estimate  the
uncertainty in the Higgs $p_\perp$-distribution,  caused by unknown higher-order QCD 
corrections to the bottom quark  contribution.

The goal of this paper is  to make a  step towards a better understanding of the
origin of  double logarithmic corrections to the Higgs boson production, their
computation  in the two-loop approximation and  to
their resummation.  Since these tasks are 
very challenging,  we 
restrict our  analysis to   {\it abelian}  QCD corrections, {\it
i.e.}  corrections associated with the abelian color factor $C_F^n$ in the $n$-th
order of QCD perturbation theory. Note that the abelian radiative corrections are
generated by the coupling of virtual gluons  to massive off-shell quarks. As a
consequence, these corrections are infra-red finite on their own, so that physical 
results can be obtained without the need to consider 
processes with additional soft and collinear radiation.

The paper is organized  as follows. In the next section we introduce our
notation. In Section~\ref{sec::1l} we describe evaluation of   one-loop
double logarithmic corrections to the bottom quark contribution to $gg \to Hg$  helicity amplitudes. In
Section~\ref{sec::2l} we extend this analysis to two loops.  In
Section~\ref{sec::resum} we show how these logarithmic corrections can be
resummed to all orders in the strong coupling constant. Numerical estimates
of the corrections are given in Section~\ref{sec::num}. We conclude in Section~\ref{sec::conc}.

\section{Setup and notations}

We consider the  Higgs boson production in the process $gg \to Hg$ mediated by the bottom quark 
loop. The Higgs boson has a non-vanishing transverse momentum. 
The particle momenta are
assigned in the following way \be
g(p_1 ) + g(p_2) \to g(p_3) + H(p_H).
\label{eq_process}
\ee
Our goal is to find the double logarithmic contributions to helicity amplitudes
in a kinematic situation where the energy of the final state gluon $E_3$  is
much smaller than the energies of the colliding gluons $E_{1,2}$ and the  Higgs
boson mass.  At the same time, we consider $E_3$ to be much  larger than the mass of the
quark that  mediates the $gg \to H$ transition.  When written in terms of
kinematic invariants, these conditions imply 
\be
m_b^2 \ll p_\perp^2 \ll  t, u  \ll s,m_H^2,
\ee
where $s = (p_1+p_2)^2, t = (p_1 - p_3)^2, u=(p_2-p_3)^2$ and $p_\perp^2 = tu/s$
is the square of the transverse momentum of the Higgs boson or the gluon in the
final state.

To illustrate this kinematic situation further, consider the production of a Higgs
boson through a bottom quark loop accompanied by an emission of a soft gluon.
We take $m_b = 4.2~{\rm GeV}$, $\sqrt{s} \approx m_H$,  $p_\perp \approx 20~{\rm
GeV}$ and assume central production (small rapidity), so that $E_3 \approx p_\perp$. Numerically we
find
\be
\frac{m_b}{E_1} \sim \lambda^{2}, \;\;\; \frac{m_b}{ E_3} \sim \lambda,
\,\;\;\; \frac{E_3}{E_1} \sim \lambda,
\label{eq::scaling}
\ee
with  $\lambda \sim 0.25$. We consider $\lambda$ to be a small parameter and
adopt the scaling rules~Eq.(\ref{eq::scaling}). In the limit $\lambda\to 0$  the
$gg\to Hg$ amplitude develops  the $1/\lambda$ singularity,  characteristic to the
soft gluon emissions; this allows us to write  the perturbative series for the amplitude in the following way
\be
{\cal M}_{gg \to Hg} =  \frac{g_s}{\lambda} \sum \limits_{n=1}^{\infty} C_{n}
\alpha_s^n \ln^{2 n}(\lambda)+\ldots\,.
\label{eq125}
\ee
In Eq.(\ref{eq125}), we neglected  all  terms that  are less singular than $\lambda^{-1} \alpha_s^{n} \ln^{2n} \lambda$ 
in the $\lambda \to 0$ limit. 
We are interested in the abelian part of the
coefficients $C_n$, which determine the double logarithmic approximation for the
amplitude. The leading-order  coefficient $C_1$ is    well-known and can be extracted
from the  one-loop result for the $gg \to Hg$ amplitude \cite{Baur:1989cm}. In what follows, we
explain how to obtain this coefficient {\it without}  following the standard route of a 
one-loop computation.  We then compute the two-loop coefficient
$C_2$ and generalize  the result to  arbitrary $n$.

We begin by fixing the notation for  helicity amplitudes. There are eight helicity 
amplitudes that are needed to describe  $g_1 g_2 \to H g_3$ process. 
However, when the
gluon $g_3$ is soft, the Higgs boson is effectively produced in the  collision of two
energetic gluons $g_1$ and $g_2$. This can only happen when 
helicities of these gluons are equal.  The constraint $\lambda_1 = \lambda_2$
leaves us with four helicity amplitudes  which  are pair-wise related by the parity
conjugation.  We take $M_{+++}$ and $M_{++-}$ as  the two 
independent amplitudes that we need to compute.

It is convenient to write the amplitudes in such a way that their  spin-helicity
structure in the soft limit is factored out, and the remaining part  only depends 
on the Mandelstam invariants of the process
\be
\begin{split}
& M_{+++}^{\rm soft} = -g_s \sqrt{2} f^{a_1 a_2 a_3} \frac{g_s^2 g_y m_b}{16 \pi^2 }
\frac{ \langle 1 2 \rangle ^2  }{ [12] \langle 2 3 \rangle \langle 1 3 \rangle } \;
A_{+++}(t,u,m_H^2,m_b^2),
\\
& M_{++-}^{\rm soft} =  - g_s \sqrt{2} f^{a_1 a_2 a_3} \frac{g_s^2 g_Y m_b}{16 \pi^2}
\frac{ \langle 12 \rangle }{[23] [13]}
\; A_{++-}(t,u,m_H^2,m_b^2).
\end{split}
\ee
Two helicity-dependent form factors $A_{++\pm}$ are given by the series in the
strong coupling constant
\be
A_{++\pm} = A_{++\pm}^{(0)} + \left ( \frac{ \alpha_s}{2 \pi} \right )
\left ( C_F A_{++\pm}^{(1A)}+C_A  A_{++\pm}^{(1NA)} \right )
 + {\cal O}(\alpha_s^2),
\label{eqAhel}
\ee
where the abelian and non-abelian parts are  separated. Our goal is to compute abelian contributions 
at two loops and beyond.

\section{One-loop helicity amplitudes in the double logarithmic approximation}
\label{sec::1l}

\begin{figure}[t]
\begin{center}
\begin{tabular}{ccc}
\includegraphics[width=4cm]{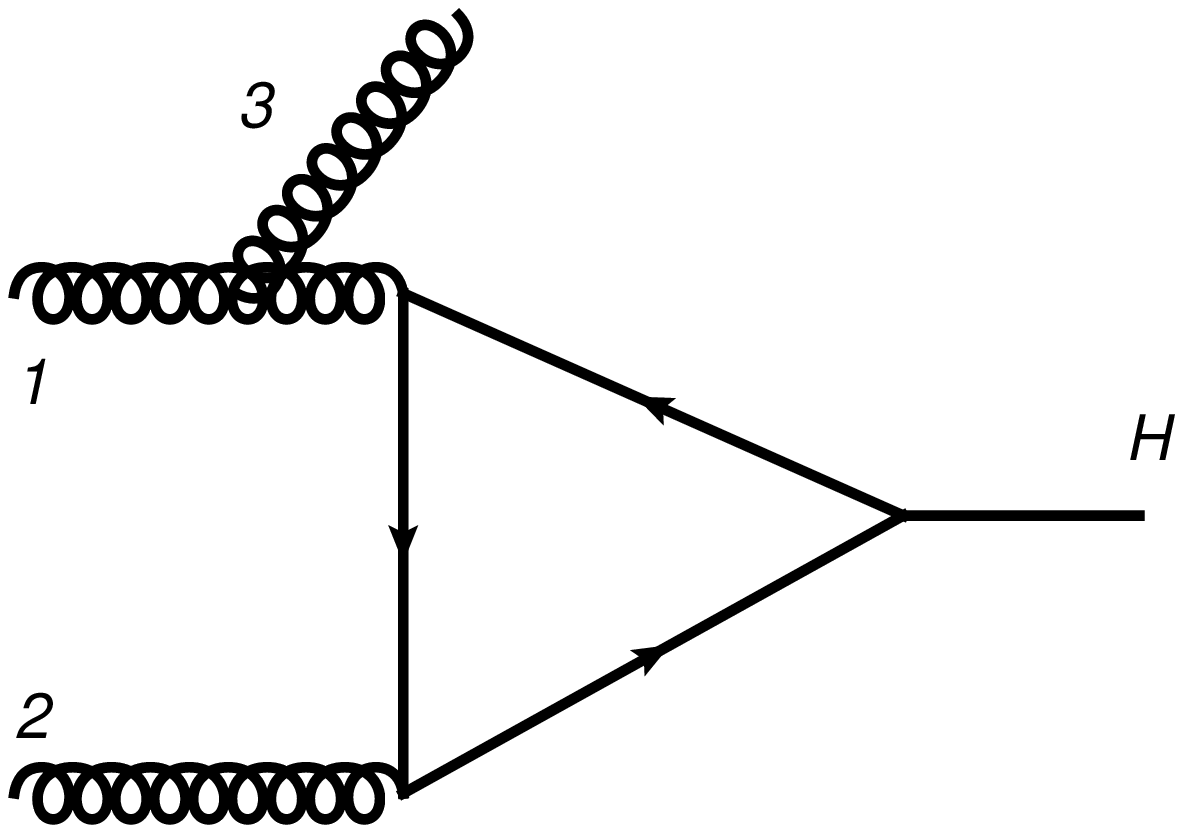}&
\includegraphics[width=4cm]{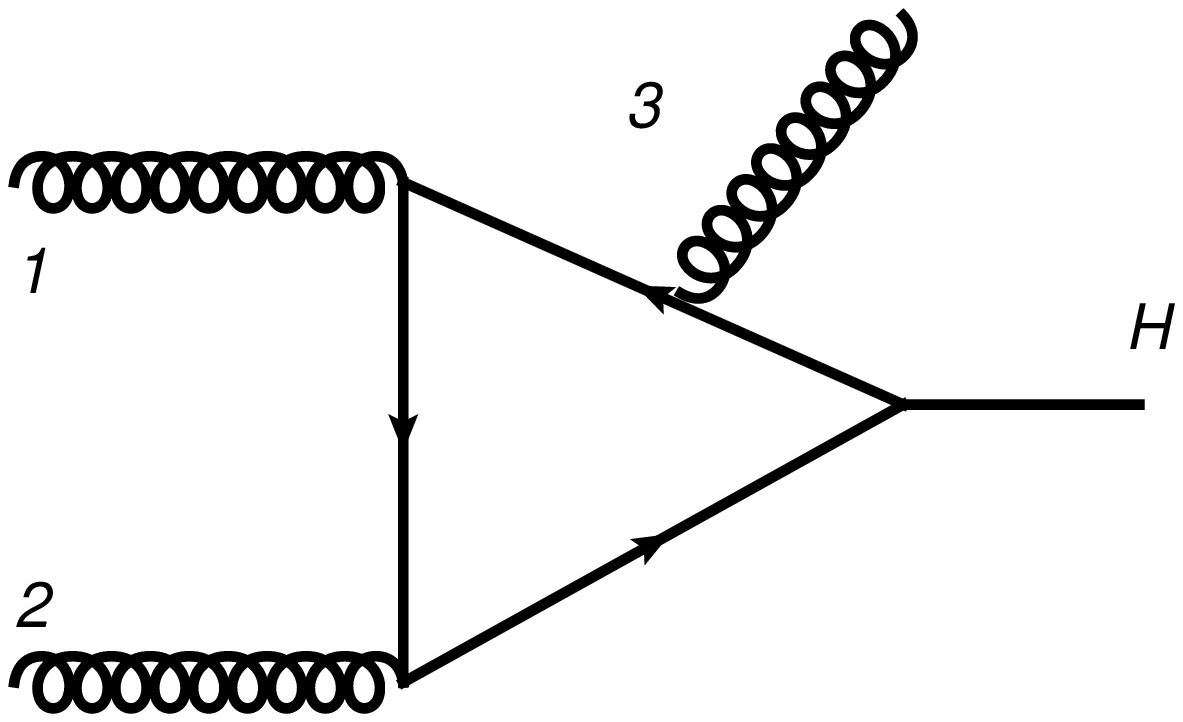}&
\includegraphics[width=4cm]{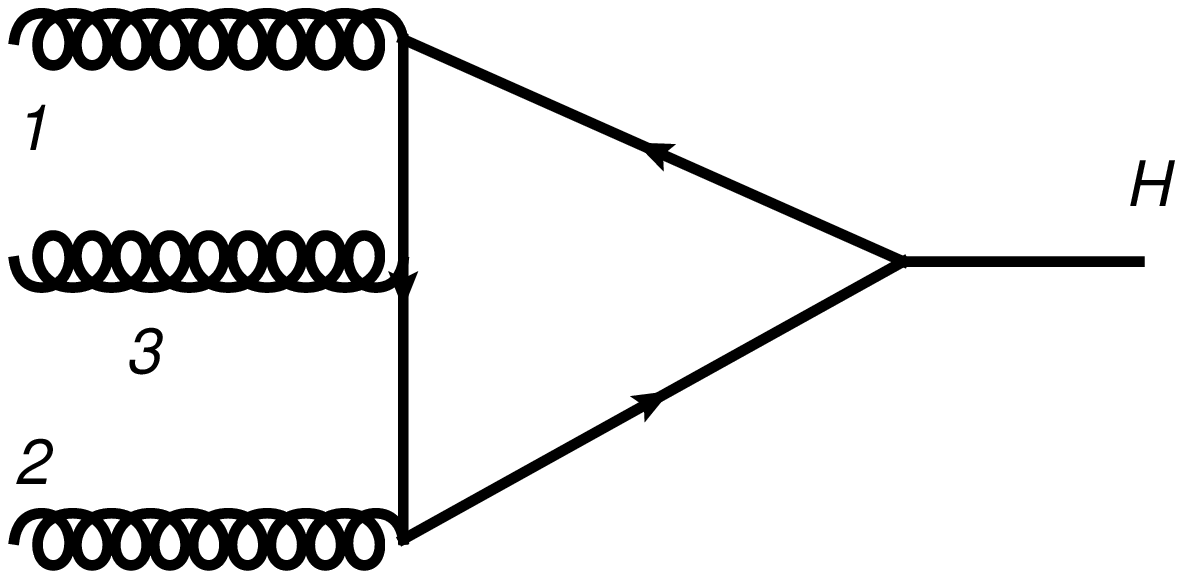}\\
&&\\
a)&b)&c)
\end{tabular}
\end{center}
\caption{\label{fig::1loop}  
One-loop diagrams representing the leading order
bottom quark contribution to $gg \to Hg$ process. Symmetric diagrams
corresponding to the opposite direction of the quark flow and to the soft
emissions off  the opposite gluon/quark line are not shown. }
\end{figure}

In this section, we will study the double logarithmic contributions to the one-loop $gg \to Hg$ 
amplitude mediated by a quark of mass $m$. A well-known example of a double logarithmic enhancement 
is provided by the Sudakov logarithms \cite{Sudakov:1954sw}. However, the situation with $gg \to Hg$ is 
different. Indeed,  as we will show, in contrast to the Sudakov logarithms \cite{Sudakov:1954sw} associated with the
radiation  of soft virtual gauge bosons  by  highly energetic on-shell
charged particles, the double logarithmic enhancement of the $gg \to Hg$ amplitude
is caused by a  soft quark exchange. Such non-Sudakov double logarithms are
typical  for amplitudes that are   mass-suppressed  at high energy
\cite{Gorshkov:1966ht,Kotsky:1997rq,Penin:2014msa}. Since physics of these  non-Sudakov double 
logarithmic corrections  is   not well-known, we begin by  discussing  the one-loop case in  detail.

In total, there are  ten  
one-loop Feynman diagrams contributing to the leading order $gg \to Hg$ 
amplitude, Fig.\ref{fig::1loop}. However,  up to differences in color factors that ensure that the final 
result is proportional to structure constants $f^{abc}$ of the gauge group  SU$(3)$,  
diagrams that  differ only by  the direction of the quark flow in the loop give identical contributions.
The number of relevant  diagrams can be  further reduced by a judicious  choice of 
gluon polarization vectors.  Indeed, each polarization vector can be chosen 
to satisfy two transversality conditions. It is convenient to require
\be
\epsilon_i \cdot p_i = 0,\;\; i \in \{1,2,3\},
\;\;\; \epsilon_1 \cdot p_2 = 0,\;\;\; \epsilon_2 \cdot p_1 = 0,\;\;\;
\epsilon_3 \cdot p_2 = 0.
\label{eq:gauge}
\ee
Explicit expressions for polarization vectors satisfying Eq.(\ref{eq:gauge})
in terms of spinor products are given in Appendix. Emission of 
a soft gluon $g_3$ off   the gluon or quark line carrying  large momentum 
$p_i$ can be described by an effective vertex  proportional to $p_i \cdot \epsilon_3$.  Thus the  condition
$\epsilon_3 \cdot p_2 =0 $ ensures that there are no soft gluon emissions  off
the gluon and quark lines carrying the external momentum $p_2$.  As the  result,
only diagrams  shown in Fig.\ref{fig::1loop} need to be 
considered.

To determine   the double logarithmic asymptotic behavior of the amplitude we
follow the original method of Ref.~\cite{Sudakov:1954sw}.  We start  by 
evaluating   the diagram Fig.\ref{fig::1loop}a  together with the diagram with 
the opposite direction of the quark flow.  By calculating the trace we 
find that the diagram  is proportional to $m_b$. The mass suppression is
caused by the fact that the  Higgs-quark  interaction flips quark helicity.
Since strong interactions conserve helicity in the massless  limit, the 
mass term provides the required second helicity flip 
in the quark loop. This helicity flip is caused  by a mass term 
of the soft quark propagator
\be
 \frac{\hat l + m_b }{l^2 - m_b^2} \to \frac{m_b}{l^2 - m_b^2}.
\label{eq3.2}
\ee
It follows from Eq.(\ref{eq3.2}) that 
once the mass term is selected, the soft quark propagator becomes a propagator
of a scalar particle, which  is sufficiently singular at small momenta to
develop a double logarithmic contribution.   We note that if the mass term is
taken  from the quark propagator that carries large momentum, the double
logarithmic contribution does not develop because the soft quark  propagator $1/\hat l$ is 
 insufficiently singular. By virtue of a similar argument,  the soft  loop momentum $l$ can often be 
neglected in  the numerators of contributing  diagrams, since  we are only interested in the leading logarithmic 
enhancement.  We  note that the last feature is not 
generic  (see e.g. the  analysis of the diagram Fig.\ref{fig::1loop}c below).
With all these simplifications it is  straightforward to derive  
contributions of the diagram Fig.\ref{fig::1loop}a to the helicity-dependent form factors. They  read
\be
A_{++\pm}^{(0),1a} = -32\; i\; \pi^2 s\; C(s,t,m_b^2),
\label{eqres1a}
\ee
where
\begin{equation}
C(s,t,m_b^2) = \int {{\rm d}^4l \over (2\pi)^4}
{1\over (l^2-m_b^2)((p_1-p_3-l)^2-m_b^2)( (p_2+l)^2-m_b^2)}
\label{eq::triangle}
\end{equation}
is the three-point function with two of its legs off-shell.  To compute
$C(s,t,m_b^2)$ in the double logarithmic approximation, we follow 
Ref.~\cite{Sudakov:1954sw} and  introduce the Sudakov parametrization of
the virtual momentum $l=\alpha p_1+\beta p_2+l_\perp$.  We  
integrate over the transverse momentum components $l_\perp$ by taking the 
residue of the soft quark propagator pole
\be
\frac{1}{l^2-m_b^2 + i0 } \to -i\pi \delta(l^2 - m_b^2) =
- i \pi \delta(s \alpha \beta - l_\perp^2 - m_b^2).
\ee
This allows for a symmetric treatment of the  soft and collinear parts of the
double logarithmic contribution. The two remaining propagators in
Eq.(\ref{eq::triangle}) become
\be
\frac{1}{(p_2 -l)^2-m_b^2}\to \frac{1}{s \alpha},\;\;\;\;\;\;
\frac{1}{(p_1 - p_3 -l)^2-m_b^2} \to \frac{1}{t - \beta s}.
\ee
To obtain the double logarithmic contribution we require both $\alpha$ and
$\beta$ integrations to be logarithmic.  This requirement is automatically
satisfied for  the integration over $\alpha$. At the same time the
integration over $\beta$ is logarithmic only for $\beta > |t|/s$. Hence, in the
double logarithmic approximation  Eq.(\ref{eq::triangle})  reduces to
\begin{equation}
C(s,t,m_b^2) \approx {i\over 16\pi^2 s}\int_{m_b^2/s}^1{{\rm d}\alpha\over \alpha}
\int^1_{|t|/s}{{\rm d}\beta\over \beta} \theta(\alpha\beta-m_b^2/s),
\label{eq3.7}
\end{equation}
where the intervals $|t|/s<\beta<1$ and $m_b^2/s<\alpha<1$  are determined by
the effective  infrared and ultraviolet cutoffs of the logarithmic integral, and
additional kinematic constraint  $\alpha\beta>m_b^2/s$  ensures that the pole
of the soft quark propagator is  in the integration domain.
It is convenient to factor out the large logarithm $L=\ln{(m_b^2/s)}\approx
\ln{(m_b^2/m_H^2)}$ and introduce the normalized  variables $\eta=\ln \alpha /L$
 and $\xi=\ln \beta/L$. By using Eqs.(\ref{eqres1a},\ref{eq::triangle},\ref{eq3.7}) we find
\begin{equation}
A_{++ \pm}^{(0),1a} =\pm 2L^2\int_0^{1-\tau_t}{\rm d}\eta\int_0^{1-\eta}
{\rm d}\xi
=\pm{L^2}(1-\tau_t^2),
\label{eq::1a}
\end{equation}
where $\tau_t=\ln\left({m_b^2/ | t| }\right)/L$.

Next, we consider  the  diagram   Fig.\ref{fig::1loop}b.  To compute this
diagram in the double logarithmic approximation, we again pick  up a mass term
from the soft quark in the $t$-channel and neglect  the momenta $l$ and $p_3$ everywhere
in the numerator. Then the contribution of the diagram Fig.\ref{fig::1loop}b 
reduces to
\be
A_{++ \pm}^{(0),1b} = -16 i \pi^2 ts D(s,t,m_H^2,m_b^2),
\ee
where $D(s,t,m_H^2,m_b^2)$ is the four-point integral
\begin{equation}
D  = \int {{\rm d}^4l \over (2\pi)^4}{1\over (l^2-m_b^2)
((p_1-l)^2-m_b^2) (p_1-p_3-l)^2-m_b^2)( (p_2+l)^2-m_b^2)}.
\label{eq::D}
\end{equation}
We use the same Sudakov parametrization $l = \alpha p_1 + \beta p_2 + l_\perp$
as before. Upon inspecting the infrared structure of Eq.(\ref{eq::D}) we find
that the double logarithmic contribution can only be obtained  when the
propagator
\be
\frac{1}{(p_1-p_3-l)^2-m_b^2} \approx \frac{1}{t - \beta s}
\ee
becomes independent of $\beta$. This leads to a constraint  $\beta< |t|/s$. The
logarithmic integration intervals become $m_b^2/s<\alpha<1$, $m_b^2/s<\beta<|t|/s$
and we obtain
\begin{equation}
A_{++ \pm}^{(0),1b}  =\pm L^2
\int_{1-\tau_t}^1{\rm d}\eta\int_0^{1-\eta}{\rm d}\xi
=\pm {L^2}{\tau_t^2\over 2}.
\label{eq::1b}
\end{equation}

We will now discuss  the diagram shown in Fig.\ref{fig::1loop}c
where the gluon is emitted off the soft quark line. Similar to the previous case
we deal here with the box diagram and  need to ``remove''  one of its propagators 
to obtain  the proper (logarithmic)  scaling of the integrand. In fact,  the underlying box diagram
has {\it two} non-overlapping momenta regions that  lead to 
a double logarithmic enhancement. These  regions are characterized by
the choice of the soft  momentum  in the diagram. Indeed, we can choose the soft momentum 
$l$ in such a way that the momentum of the emitted gluon  $p_3$  flows through  the lower (upper)
half of the quark loop Fig.\ref{fig::1loop}c in region I (II), respectively. Consider 
region I and choose the momentum decomposition $l = \alpha p_1 + \beta p_3 +
l_\perp$.  After omitting irrelevant terms, the quark propagators become
\begin{eqnarray}
&&\frac{\hat l + m_b }{l^2 - m_b^2} \to - i\pi m_b \delta ( |t| \alpha \beta - l_\perp^2 - m_b^2),
\label{eq::prop3}
\\
&&\frac{  \hat p_1 - \hat l+ m_b
}{(p_1-l)^2 - m_b^2}  \to  \frac{ \hat p_1 }{t \beta},
\label{eq::prop4}
\\
&&\frac{\hat  p_3 -\hat l+ m_b }{(p_3-l)^2 - m_b^2} \to
\frac{   \hat p_3 -\alpha \hat p_1  }{t \alpha},
\label{eq::prop1}
\\
&&\frac{ \hat p_2 + \hat p_3 -  \hat l + m_b
}{(p_2 - p_3 +l)^2 - m_b^2} \to  \frac{ \hat p_2 }{u + s  \alpha}.
\label{eq::prop2}
\end{eqnarray}
It follows from Eqs.(\ref{eq::prop3},\ref{eq::prop4},\ref{eq::prop1},\ref{eq::prop2}) 
that the double logarithmic contribution can be obtained in two different ways:
(i) for  $ \alpha  < |u|/s$ only the $p_3$ term in the numerator  of
Eq.(\ref{eq::prop1}) should be kept (the ``scalar'' contribution)  and (ii)
for  $\alpha > |u|/s$ the $\alpha p_1$ term  should be taken  from the
numerator  in Eq.(\ref{eq::prop1})
to cancel  an extra power of $\alpha$ that appears in the denominator
of Eq.(\ref{eq::prop2}) in this limit (the ``vector'' contribution).\footnote{ 
We refer to this contribution as ``vector'' because  it originates  
from a term in the numerator of the diagram Fig.\ref{fig::1loop}c which is linear 
in the soft loop momentum $l$.}  These 
two contributions are proportional to
\begin{equation}
N^s_{\lambda_1,\lambda_2,\lambda_3}  = {{\rm Tr} \left[
\hat \epsilon_3 \hat \epsilon_1 \hat p_1 \hat p_2 \hat \epsilon_2 \hat p_3
\right ]\over 2tu},
\end{equation}
and
\begin{equation}
N^v_{\lambda_1,\lambda_2,\lambda_3} = {{\rm Tr} \left[
\hat \epsilon_3 \hat \epsilon_1 \hat p_1 \hat p_2 \hat \epsilon_2 \hat p_1
\right ]\over 2ts},
\end{equation}
respectively. By calculating traces and using  explicit expressions
for the polarization vectors given in Appendix, we find the  following results
\be
\begin{split}
& N^s_{+,+,+} =0,\;\;\;\;\;\;\;\;\;\;\;\;\;\;\;\;\;\;\;\;\;\;\;
N^s_{++-} = \frac{\sqrt{2} \langle 1 2 \rangle }{[13][23]}, \\
& N^v_{+,+,+} = \frac{ \sqrt{2} \langle 12 \rangle ^2 }{ [ 12 ] \langle 2 3 \rangle \langle 1 3 \rangle},
\;\;\;\;\;
N^v_{++-} = -\frac{\sqrt{2} \langle 1 2 \rangle }{[13][23]},
\end{split}
\ee
Note that the vector integral has the usual tensor structure of a color-dipole
emission  which gives  $A_{+++} = - A_{++-}$, similar to all  other diagrams. At the
same time,   the tensor structure of the scalar integral corresponds to the
three-gluon configuration  described by a local gauge invariant operator
\begin{equation}
G_{\mu\nu}^aG^{b}_{\nu\lambda}G_{\lambda\mu}^cf^{abc},
\label{eq::Gcube}
\end{equation}
which  does not contribute   to the all-plus  helicity amplitude.

By crossing symmetry,  the scalar contributions of the momentum  regions I and
II are equal. Therefore,  the total scalar contribution of the  diagram
Fig.\ref{fig::1loop}c  can be written in terms of the  double logarithmic
integral over the interval  $m_b^2/|t|<\alpha<|u|/s$, $m_b^2/|t|<\beta<1$ that originates  from  
region I. This  gives
\begin{equation}
A^{(0),1c,s}_{++-} =-2L^2\int_{1-\tau_u}^{\tau_t}{\rm d}\eta\int_0^{\tau_t-\eta}{\rm d}\xi
=-L^2{(1-\tau_t-\tau_u)^2},
\label{eq3.18}
\end{equation}
whereas $A^{(0),1c,s}_{+++}=0$. At the same time,   the vector contribution of 
region II vanishes due to our choice of the polarization vector for the gluon 
$g_3$,  $p_2 \cdot \ep_3 = 0$. As the result,   the  total vector contribution of the  diagram
Fig.\ref{fig::1loop}c is given by the  double logarithmic integral over the
interval  $|u|/s<\alpha<1$, $m_b^2/|t|<\beta<1$ from region I. It  reads
\begin{equation}
A^{(0),1c,v}_{++ \pm} =\pm L^2
\int_0^{1-\tau_u}{\rm d}\eta\int_0^{\tau_t-\eta}{\rm d}\xi
=\mp{L^2}{(1-\tau_u)(1-2\tau_t-\tau_u)\over 2}.
\label{eq::1c}
\end{equation}
We are now in position to present the leading-order bottom-quark contribution 
to  $gg \to Hg$ helicity amplitudes in the double logarithmic approximation. 
We sum the contributions of individual diagrams  given in 
Eqs.(\ref{eq::1a},\ref{eq::1b},\ref{eq3.18},\ref{eq::1c}) and obtain
\be
\begin{split}
& A_{+++}^{(0)} = L^2\left(1- \frac{\tau^2}{2} \right),\;\;\;\;\;A_{++-}^{(0)}
= -L^2\left( 1 +  \frac{\tau^2}{2} \right),
\label{eqprev}
\end{split}
\ee
where we used $\tau=\ln(m_b^2/p^2_\perp)/L$. These results coincide with the 
double logarithmic limits of the one-loop amplitudes computed in Ref.~\cite{Baur:1989cm} 
long time ago.\footnote{See also Ref.~\cite{Banfi:2013eda} for a recent discussion.}
Our analysis identifies  the origin of  the double logarithmic 
enhancement of  the $gg \to Hg$ amplitude mediated by a light quark.  
With this understanding, it is straightforward to extend the above calculation first to  
two loops and then to all orders in the strong coupling constant $\alpha_s$. 
We will describe how to do this in the next sections.

\section{Two-loop helicity amplitudes in the double logarithmic approximation}
\label{sec::2l}

\begin{figure}[t]
\begin{center}
\begin{tabular}{ccc}
\includegraphics[width=4cm]{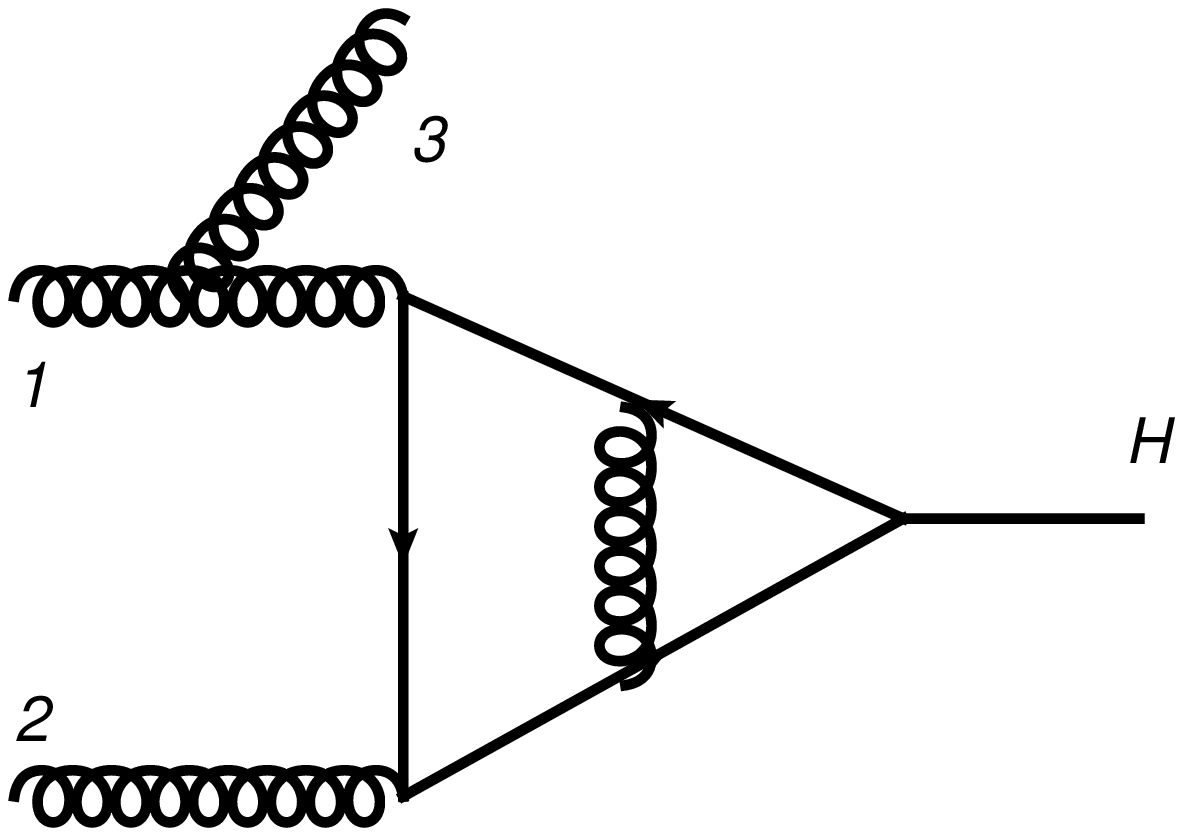}&
\includegraphics[width=4cm]{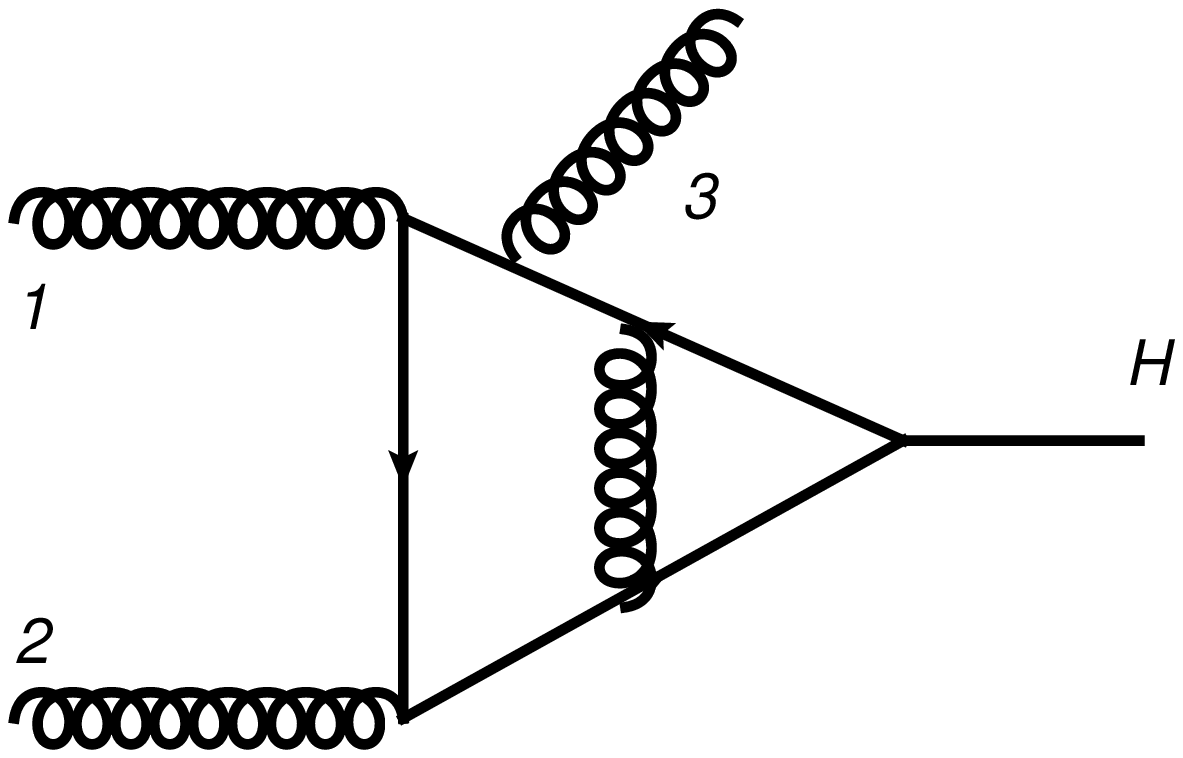}&
\includegraphics[width=4cm]{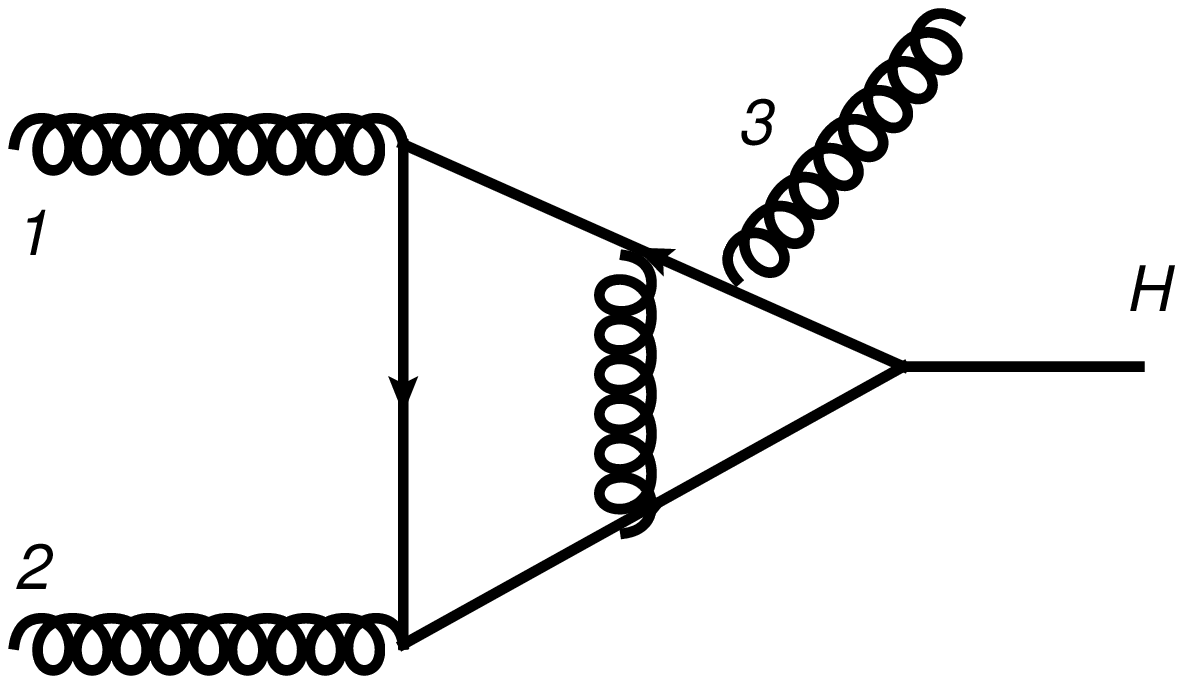}\\
&&\\
a)&b)&c)\\
&&\\
\includegraphics[width=4cm]{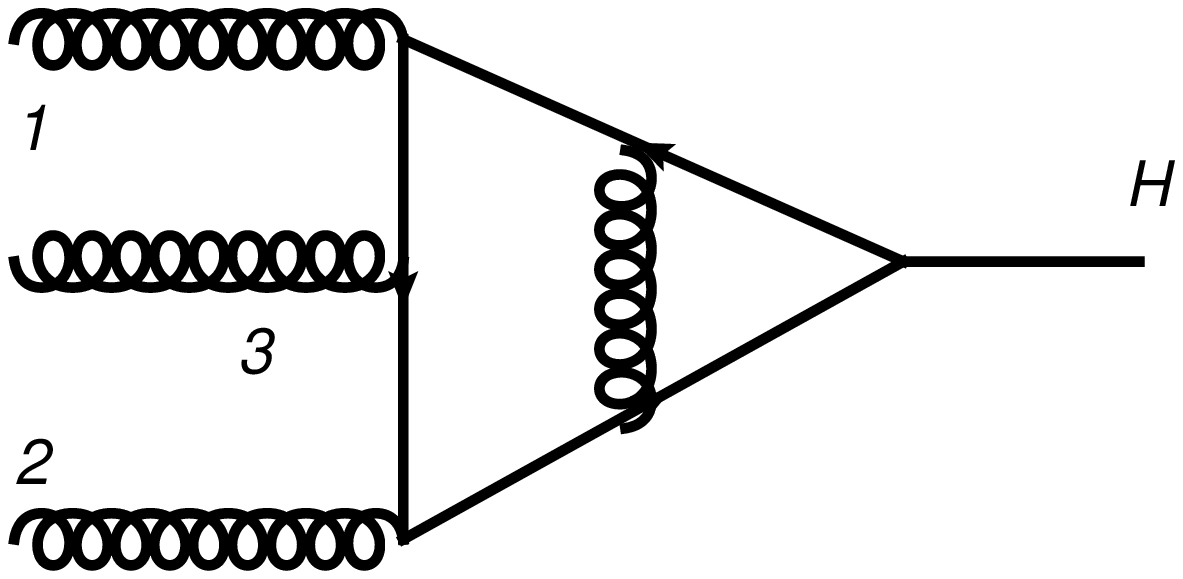}&
\includegraphics[width=4cm]{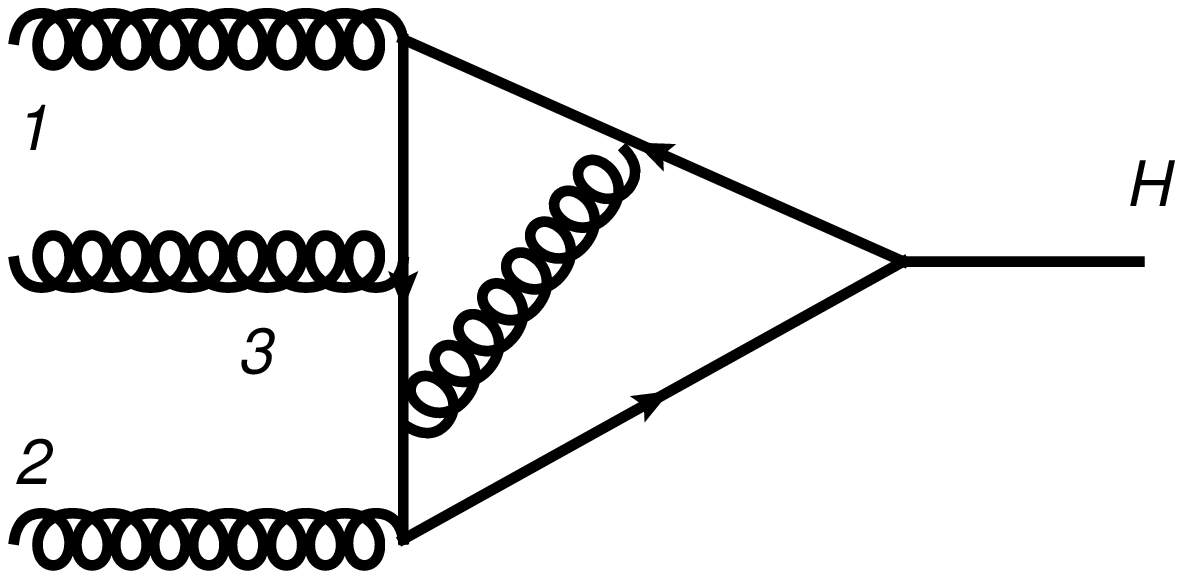}&
\includegraphics[width=4cm]{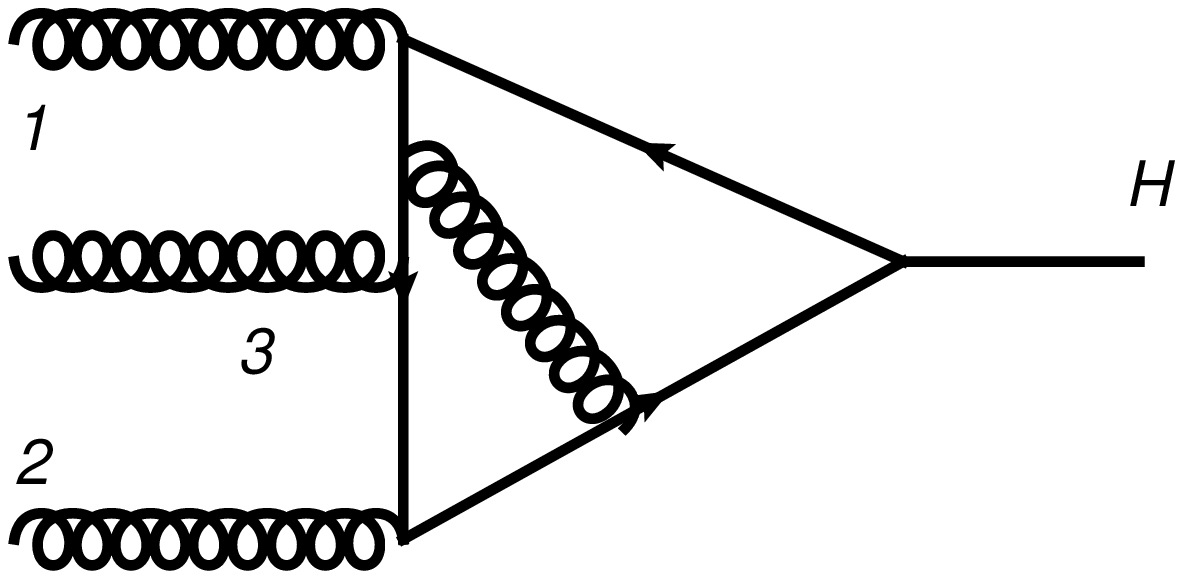}\\
&&\\
d)&e)&f)
\end{tabular}
\end{center}
\caption{\label{fig::2loop}  Two-loop  diagrams  contributing to the
abelian double logarithmic corrections. Diagrams that differ by the direction of
the fermion flow are not shown.
}
\end{figure}

It is easy to convince oneself that a two-loop diagram contributing to $gg \to Hg$ 
can develop  leading ${\cal O}(m_b)$ double logarithmic enhancement 
only if exactly one of its fermion lines is soft. 
Indeed,  since each  soft fermion  effectively  contributes  one power of $m_b$ to the final 
result,  leading ${\cal O}(m_b)$ double logarithms  
are provided by exchanges of   one soft fermion and one soft virtual gluon.  

The abelian part of the two-loop correction originates from  a soft
gluon exchange  between   virtual bottom quarks. The relevant two-loop Feynman
diagrams are shown  in  Fig.\ref{fig::2loop}. We note that the one-loop correction 
to $q \bar q H$ vertex appears as a subdiagram in many two-loop diagrams 
in  Fig.\ref{fig::2loop}. This correction develops a double logarithmic enhancement, 
so that the (properly normalized)  $q \bar q H$ vertex in the one-loop approximation reads
\cite{Sudakov:1954sw}
\begin{equation}
V_{q  \bar q H} = 1 + \delta V_{q \bar q H},
\;\;\;\;\;\;\;\;\;
\delta V_{q \bar qH} = 
-{C_F\alpha_s\over 2\pi}\ln\left({q_1^2\over 2q_1 \cdot q_2 }\right)
\ln\left({q_2^2\over 2q_1 \cdot q_2 }\right).
\label{eq::factor}
\end{equation}
In Eq.(\ref{eq::factor}), $q_1$ and $q_2$ are the momenta of the off-shell quark lines and we assume that 
$m_b^2\ll q_1^2,~q_2^2\ll q_1\cdot q_2$. This expression and the one-loop
analysis of  the previous section can be used to easily compute the leading
logarithmic part of the relevant  two-loop diagrams.

We begin with the diagram Fig.\ref{fig::2loop}a.   The external momenta of
the vertex subgraph  in this case are $q_1 = p_1-l $ and $q_2 = p_2+l$, where
$l$ is the soft momentum of the quark loop. For   $l = \alpha p_1 + \beta p_2 +
l_\perp$ we get  $q_1^2 = s\beta$, $q_2^2 = s \alpha$, $ 2 q_1 \cdot q_2 \approx
s$, so that
\begin{equation}
\delta V_{q  \bar q H}^{2a} = -{C_F \alpha_s\over 2\pi}\ln\alpha\ln\beta=-x\eta\xi,
\label{eq::factor2a}
\end{equation}
where $x=C_F\alpha_s L^2/ 2\pi$ and $\eta,\xi$ and $L$ are defined in the
previous section. The double logarithmic integration over the  soft quark
momentum is the same as for the diagram Fig.\ref{fig::1loop}a and the
correction to the  helicity amplitudes is obtained by including $\delta V_{q \bar q H}^{2a}$ factor
into the integrand of the one-loop expression  Eq.(\ref{eq::1a}). For the
two-loop abelian  coefficient in Eq.(\ref{eqAhel}) we obtain
\begin{equation}
A^{(1A),2a}_{++\pm} =\mp 2 L^4\int_0^{1-\tau_t}{\rm d}\eta\int_0^{1-\eta}
{\rm d}\xi\, \eta \; \xi =
\mp {L^4}{(1- 4\tau_t^3 +3 \tau_t^4) \over 12}.
\label{eq:2loopa}
\end{equation} The diagram in Fig.\ref{fig::2loop}b is computed in a similar
way. Virtualities of the quark lines become  $q_1^2 \approx |t|$ and $q_2^2
\sim s \alpha$, and the one-loop vertex reads 
\be
\delta V_{q  \bar q H}^{2b}=-{C_F \alpha_s\over 2\pi} \ln { |t|\over s  } \ln\alpha =-x (1 - \tau_t) \xi.
\label{eq::factor2b}
\ee
Substituting this result into Eq.(\ref{eq::1b}), we obtain
\begin{equation}
A^{(1A),2b}_{++\pm}
=\mp  L^4\int_{1-\tau_t}^1{\rm d}\eta\int_0^{1-\eta}{\rm d}\xi
\left(1-\tau_t\right)\xi
=\mp {L^4}{(\tau_t^3- \tau_t^4 ) \over 6}.
\label{eq:2loopb}
\end{equation}
The double logarithmic contribution of the diagram Fig.\ref{fig::2loop}c is
generated when the quark propagator between the emission vertex of the soft
gluon $g_3$ and the $q \bar q H$ vertex becomes independent of the soft photon
loop momenta. In this case the inner loop   reduces to the one-loop vertex
with an additional restriction  on the integration region. We find
\be
\delta V_{q \bar q H}^{2c}=-{C_F \alpha_s\over 2\pi} \ln {\beta s\over |t| } \ln\alpha
=-x \left(\eta-1+\tau_t\right)\xi .
\label{eq::factor2c}
\ee
The corresponding two-loop corrections to the amplitudes read
\begin{equation}
A^{(1A),2c}_{++\pm} =\mp L^4\int_{1-\tau_t}^1{\rm d}\eta\int_0^{1-\eta}{\rm d}\xi
\left(\eta-1+\tau_t\right)\xi
=\mp {L^4}{\tau_t^4\over 24}.
\label{eq:2loopc}
\end{equation}
To compute the diagram Fig.\ref{fig::2loop}d in the double logarithmic
approximation, we again insert the  expression for  the $q \bar q H$ 
vertex Eq.(\ref{eq::factor}) into the one-loop diagram in Fig.\ref{fig::1loop}c. As
was explained in the previous section this diagram receives the double logarithmic
contributions from two independent momentum regions.   In region I we 
parametrize the soft momentum as   $l = \alpha p_1 + \beta p_3 + l_\perp$ and
find
\be
\delta V^{2d}= -{C_F\alpha_s\over 2\pi}  \ln\frac{|t| \beta}{s} \ln  \frac{|u| + \alpha s }{s}
\to
-{C_F\alpha_s\over 2\pi}   \left \{
\begin{array}{cc}
\ln  \frac{|t| \beta}{s}  \ln  \frac{|u|}{s }\;\;\;\; & \alpha< \frac{|u|}{s }, \\
 & \\
\ln \frac{|t| \beta}{s}  \ln \alpha \;\;\;\;  & \alpha > \frac{|u|}{s }.
\end{array}
\right.
\label{eq:regionI}
\ee
We note that the two integration regions, $\alpha < |u|/s $ and  $\alpha > |u|/s $, 
correspond to   scalar and vector contributions, respectively. The expression for 
the $q \bar q H$ vertex in region II can be found in the same way. We insert 
these results into Eqs.(\ref{eq3.18},\ref{eq::1c}) and obtain
\be
\begin{split}
A_{++-}^{(1A),2d,s} & =  L^4
\int_{1-\tau_u}^{\tau_t}{\rm d}\eta\int_0^{\tau_t-\eta}{\rm d}\xi
(1-\tau_u)\left(\xi+1-\tau_t\right)
+(t\leftrightarrow u) \\
&=
{L^4}{(2-2\tau_t+\tau_u)(1-\tau_u)(1-\tau_t-\tau_u)^2\over 6}
+ (t\leftrightarrow u).
\end{split}
\label{eq:1234}
\ee
\be
\begin{split}
A_{++\pm}^{(1A),2d,v} &= \mp L^4\int_0^{1-\tau_u}{\rm d}\eta\int_0^{\tau_t-\eta}{\rm d}\xi
(\xi+1-\tau_t)\eta  \\
& = \pm {L^4}{(1-\tau_u)^2(5-12\tau_t+6\tau_t^2-2\tau_u-3\tau_u^2)\over 24}.
\end{split}
\label{eq:1235}
\ee
We note that the corresponding one-loop expression given in Eq.(\ref{eq3.18}) 
includes equal scalar contributions from regions I
and II. At  two loops, contributions of 
regions I and II  are not equal anymore; 
we separate them  in Eq.(\ref{eq:1234}) and indicate 
contribution of the region I by the corresponding integral and the contribution of 
region II by the  $t \leftrightarrow u$
symmetric term. The vector contribution  Eq.(\ref{eq:1235}) comes entirely  from 
region I, as in the one-loop case.

Diagrams shown in Fig.\ref{fig::2loop}e and Fig.\ref{fig::2loop}f are related by
crossing symmetry and  we only  consider  the evaluation  of the former. This
diagram receives the scalar contribution  from region I; 
the double logarithmic term is generated when the propagator between the emission vertex of
the gluon $g_2$ and the $q \bar q H$ annihilation vertex becomes independent on
the soft momenta. Thus as in the case of the diagram  Fig.\ref{fig::2loop}c
the inner loop reduces to the one-loop vertex integral with an additional 
restriction  on
the integration region. The effective vertex in this case reads
\begin{equation}
\delta V_{q \bar q H}^{2e} = -{C_F \alpha_s\over 2\pi}\ln{\alpha s\over u}\ln\beta
=-x\left(\eta-1+\tau_u\right)\xi.
\end{equation}
Since the  scalar contribution to the all-plus helicity amplitude vanishes, we
obtain
\be
\begin{split}
A_{++-}^{(1A),2e}=  &  L^4
\int_{1-\tau_u}^{\tau_t}{\rm d}\eta\int_0^{\tau_t-\eta}{\rm d}\xi
\left(\eta-1+\tau_u\right)\xi
 = {L^4}{(1-\tau_t-\tau_u)^4\over 24}.
\end{split}
\label{eq::2loope}
\ee
The result for the diagram in Fig.\ref{fig::2loop}f is also given by 
Eq.({\ref{eq::2loope}}) since it is symmetric with respect to the replacement $t
\leftrightarrow u$.

Taking the sum of all the individual contributions in 
Eqs.(\ref{eq:2loopa},\ref{eq:2loopb},\ref{eq:2loopc},\ref{eq:1234},\ref{eq:1235},\ref{eq::2loope}),
we obtain the two-loop correction to the $gg \to Hg$ amplitude in the 
double logarithmic approximation
\be
\begin{split}
& A_{+++}^{(1A)} =-{L^2\over 24}\left(2-{3\tau^2}+{2\tau^3}+{3\tau^2\zeta^2}\right),
\\
& A_{++-}^{(1A)} ={L^2\over 24}\left(2+{3\tau^2}-{6\tau^3}+{4\tau^4}-{3\tau^2\zeta^2}\right),
\end{split}
\ee
where the new variable   $\zeta=\ln(u/t)/L$ parametrizes the dependence of the amplitudes 
on the soft gluon rapidity.

\section{Resummation of the abelian double logarithmic contributions}
\label{sec::resum}
The perturbative expansion parameter for the double logarithmic corrections 
$ x = \frac{C_F \alpha_s}{2\pi} L^2$ is not   small numerically,  $x \sim 1$. For this 
 reason,  resummation of such corrections might be relevant. This problem is also quite
interesting theoretically, since  very little is known
about the all-order structure of the power-suppressed  non-Sudakov  logarithms.
Indeed, on the one hand, only few examples of the
resummation of non-Sudakov double logarithmic corrections   are  known so far
\cite{Gorshkov:1966ht,Kotsky:1997rq,Penin:2014msa} and, on the other hand, 
systematic renormalization group analysis of the high-energy behavior of the
on-shell amplitudes beyond  the leading-power approximation is still elusive for
 existing effective field theory methods. 

The problem that we discuss in this paper is, however, simpler than the general case. 
As  we  pointed out already,    to leading order in $m_b$,  
higher-order double logarithmic corrections to the 
helicity amplitudes are caused by multiple  soft virtual  gluon exchanges  and  
a single soft quark exchange.   Thus we have to
consider Feynman diagrams similar to Fig.\ref{fig::2loop} but with 
multiple soft gluon  exchanges between different  quark lines. For the
abelian part of the corrections we can use simple factorization properties
of soft emissions in QED. It is well-known that 
in this case, upon summing over all relevant diagrams,  integrations over soft gluon
momenta  factorize  and the all-order result  is given by the exponent of the 
single gluon contribution,  given by the ${\cal O}(\alpha_s)$ term in Eq.(\ref{eq::factor}).

By using the expression Eq.(\ref{eq::factor2a}) specific for  the diagram
Fig.\ref{fig::2loop}a  we find the Sudakov exponent  to be
$e^{-x\xi\eta}$. The  all-order double logarithmic corrections to helicity
amplitudes are then obtained by including this exponent into the integrand of
Eq.(\ref{eq::1a}). Upon  integration   over $\xi$,   we obtain  the resummed
expression for helicity amplitudes in the form of the one-parameter integral
\begin{equation}
A_{++\pm}^{\rm A,a}  =\pm 2L^2\int_{0}^{1-\tau_t}
{1-e^{-x\eta(1-\eta)}\over x\eta}{\rm d}\eta.
\label{eq::resummeda}
\end{equation}
The multiple  gluon exchange diagrams related to 
Fig.\ref{fig::2loop}b and Fig.\ref{fig::2loop}c must  be considered
simultaneously.\footnote{Indeed, already at the two-loop level, 
these diagrams describe the two possible ways to emit the soft gluon 
with momentum $p_3$  and a soft virtual gluon by an energetic quark line.}
 After summing over all possible permutations of the soft gluon
emission vertices, their contributions factorize and produce  a {\it product} of
the exponents of the one-loop contributions~(\ref{eq::factor2b},
\ref{eq::factor2c}). They combine into the  exponential factor  $e^{-x\xi\eta}$ 
identical to the previous case. By including it into the integrand of the one-loop 
expression Eq.(\ref{eq::1b}) we get the all-order result
\begin{equation}
A_{++\pm}^{\rm A,bc}
=\pm L^2\int_{1-\tau_t}^{1}
{1-e^{-x\eta(1-\eta)}\over x\eta}{\rm d}\eta .
\label{eq::resummedbc}
\end{equation}
For the diagram Fig.\ref{fig::2loop}d the Sudakov factor depends on whether a
vector or a scalar contribution is considered, {\it cf.} Eq.(\ref{eq:regionI}). 
For the vector part,  the Sudakov exponent is $e^{-x(\xi+1-\tau_t)\eta}$ and the
all-order result associated with the leading-order contribution Eq.(\ref{eq::1c})
reads
\begin{equation}
A_{++\pm}^{\rm A,d,v} =\pm L^2\int_{0}^{1-\tau_u}{e^{-x(1-\tau_t)\eta}
-e^{-x\eta(1-\eta)}\over x\eta}{\rm d}\eta .
\label{eq::resummedd}
\end{equation}
As in the case of the diagrams Fig.\ref{fig::2loop}b and
Fig.\ref{fig::2loop}c, the  scalar contributions from the momentum region I of
the diagram Fig.\ref{fig::2loop}d combine with the  diagram
Fig.\ref{fig::2loop}e, exponentiate and produce a Sudakov factor
$e^{-x\left(1-\tau_t-\tau_u+\tau_t\tau_u+\eta\xi\right)}$. The Sudakov exponent
of the scalar contribution from  region II of the diagrams that 
combine Fig.\ref{fig::2loop}d and  Fig.\ref{fig::2loop}e with additional soft exchanges 
is obtained by the replacement $t \leftrightarrow u$.
The sum of these contributions  is, therefore, given by 
\begin{equation}
A_{++-}^{\rm A,de,s}
=-2L^2\int_{1-\tau_u}^{\tau_t}{e^{-x(1-\tau_u)(1-\tau_t)}
\left(1-e^{-x\eta (\tau_t-\eta)}\right)
\over x\eta}{\rm d}\eta + (t\leftrightarrow u).
\label{eq::resummedde}
\end{equation}
The sum of individual contributions 
given in Eqs.(\ref{eq::resummeda},\ref{eq::resummedbc},\ref{eq::resummedd},\ref{eq::resummedde}) 
determines  the complete result for the abelian double logarithmic corrections 
to the bottom quark contribution to $gg \to Hg$  helicity amplitudes to all orders in QCD  perturbation theory.

\section{Double logarithmic  corrections to the differential cross section}
\label{sec::num}
We are now in  position to estimate the effect of the corrections, computed in the previous 
section,  on the differential cross section of the Higgs boson
production in association with a jet. The total amplitude of this process is
given by the sum of top and bottom contributions since contributions of
lighter  quarks  are negligible. We therefore write
\be
\begin{split}
& M_{+++}^{\rm soft} = -g_s \sqrt{2} f^{a_1 a_2 a_3} \frac{g_s^2 }{16 \pi^2 v }
\frac{ \langle 1 2 \rangle ^2  }{ [12] \langle 2 3 \rangle \langle 1 3 \rangle }
\left [
A_{+++}^{(t)}
+ \frac{m_b^2}{m_H^2} A_{+++}^{(b)}
\right ],
\\
& M_{++-}^{\rm soft} =  - g_s \sqrt{2} f^{a_1 a_2 a_3} \frac{g_s^2}{16 \pi^2 v}
\frac{ \langle 12 \rangle }{[23] [13]}
\left [
 A_{++-}^{(t)}
+ \frac{m_b^2}{m_H^2} A_{++-}^{(b)}
\right].
\end{split}
\ee
Thanks to its large Yukawa coupling, the  top quark provides the dominant contribution to
the scattering amplitude. In the soft limit the real  emission from inside the top-quark 
loop is power-suppressed {\it i.e.} the soft emission factorizes with respect to the 
$gg\to H$ amplitude. The result for this contribution is well known and in the 
limit of an infinitely heavy top quark reads
\be
A_{++\pm}^{(t)} = \pm \frac{2}{3}.
\ee
There are ${\cal O}(\alpha_s)$ corrections to this formula that, however, are not 
essential for us. 

The largest effect of the bottom quark on the differential cross section is
caused by its interference with the top quark contribution. We find
\begin{equation}
d\sigma_{gg\to Hg}=
d\sigma^{(0)}_{gg\to Hg} \times \left[1-{3\over 2}{m_b^2\over m_H^2}
\left(A_{+++}^{(b)}-A_{++-}^{(b)} \right)
+{\cal O}(m_b^4)\right],
\end{equation}
where $d\sigma^{(0)}_{gg\to Hg}$ is the top quark mediated cross section, and we
neglect the finite top mass effects in the interference term. Note that since
the leading bottom quark effect is due to the interference with the top quark  mediated
amplitude, to leading order in $1/m_t$, any additional real emission contribution involves the three-gluon interaction and
does not contribute to the abelian part of the correction.   

We can now use the
result derived in the previous section for  numerical estimates. It is convenient
to express the correction to the cross section 
 through the variables $\tau = \ln (m_b^2/p_\perp^2)/L$  and $\zeta = \ln(u/t)/L$, 
  which parameterize
the dependence of the cross section on the transverse momentum and rapidity. We
obtain
\begin{equation}
{d\sigma_{gg\to Hg}}= d\sigma^{(0)}_{gg\to Hg} \times 
\left[1-{3\over 2}{m_b^2\over M_H^2}L^2 f(x,\tau,\zeta)
+{\cal O}(m_b^4)\right],
\label{eq::sigma}
\end{equation}
where
\be
\begin{split}
x f(x,\tau,\zeta) & =
\int_{0}^{1} \frac{{\rm d}\eta}{\eta} \left [ (1-e^{-x\eta(1-\eta)} )   \left (  1
+ 2 \theta\left (  1 - \tau -\zeta  - 2 \eta \right )  \right )
-  (1-e^{-x\eta \delta(\tau,\zeta)} ) \right ]
\\
&
+  e^{-x \delta(\tau,\zeta)} \int \limits_{(1-\tau+\zeta)/2}^{(1+\tau+\zeta)/2}
\frac{{\rm d}\eta}{\eta}
\left (1-e^{-x\eta(1+\tau+\zeta-2\eta)/2}\right)
+ (\zeta \to -\zeta),
\end{split}
\ee
and $\delta(\tau,\zeta) = ( (1-\tau)^2 - \zeta^2)/4$. The perturbative expansion
of the function $f$ reads
\be
f=2-{x\over 6}\left(1-\tau^3+\tau^4\right)+
{x^2\over 24}\left({4\over 15}-{\tau^3}+{2\tau^4}-
{7\tau^5\over 5}+{2\tau^6\over 5}+{\zeta^2}\left(\tau^3-\tau^4\right)\right)
+\dots. ,
\label{eq::fseries}
\ee
where ellipsis stands for terms suppressed by higher powers of $x$.

We can use the result Eq.(\ref{eq::fseries}) to estimate the impact of the QCD corrections to 
bottom quark contributions to $gg \to Hg$  on the Higgs boson 
transverse momentum distribution.   In principle, we should convolute  the partonic cross section Eq.(\ref{eq::sigma}) with 
the parton  distribution functions. However, we will now argue that, given 
the structure of the corrections shown in Eq.(\ref{eq::fseries}), 
this is not necessary. 
Indeed,   within the  accuracy of our approximation  $L
= \ln(s/m_b^2) \approx \ln (m_H^2/m_b^2)$ can be considered independent 
of the partonic center-of-mass energy.
In addition,  series in 
Eq.(\ref{eq::fseries})  shows very weak dependence on the rapidity of the soft gluon. 
Indeed,   the function $f$ in Eq.(\ref{eq::fseries})  does not depend on the 
gluon rapidity up to ${\cal O}(x)$.
Moreover,   at ${\cal O}(x^2)$ the rapidity-dependent part of the coefficient
includes only high powers of $\tau$.  If  the soft gluon is  
emitted at 
large rapidity,  
$|\zeta| \approx 1$ and  
$\tau\ll 1$. On the contrary, central emission 
with the large transverse momentum implies $|\zeta| \ll 1$ and $\tau \gg 1$.
Therefore, the the  rapidity-dependent
term is small everywhere  and can be neglected.    After  these
modifications the function $f$ depends only on the transverse momentum of the
emitted gluon or the Higgs boson.  As a result it remains unaffected by the
integration over parton distribution functions if the transverse momentum of the
Higgs boson is kept  fixed. Therefore, we can write
\be
\begin{split}
{d\sigma_{pp \to H+j}\over dp_\perp^2}
= & {d\sigma^{(0)}_{pp \to H+j}\over dp_\perp^2}
\left \{ 1 - \frac{3 m_b^2}{m_H^2} L_{\rm eff}^2  \left [
1-{x_{\rm eff} \over 12}\left(1-\tau^3+\tau^4\right)
\right. \right.  \\
& \left. \left. +
{x_{\rm eff}^2\over 48}\left({4\over 15}-{\tau^3}+{2\tau^4}-
{7\tau^5\over 5}+{2\tau^6\over 5} \right )+ {\cal O}(x^3)
\right ] + {\cal O}(m_b^4) \right \} ,
\end{split}
\label{eq:higgspt}
\ee
where $L_{\rm eff} = \ln (m_H^2/m_b^2)$ and $x_{\rm eff} = \frac{\alpha_s C_F}{2\pi}L_{\rm eff}^2$. 
We emphasize that Eq.(\ref{eq:higgspt}) only applies  to the
contribution of  $gg$ partonic channel to the production of the Higgs boson in
proton collisions and that only abelian corrections are taken into account
there.

We note that the series in Eq.(\ref{eq:higgspt}) has  peculiar structure. Indeed, 
the one-loop double logarithmic 
correction to ${\rm d} \sigma/{\rm d}p_\perp$ is {\it independent} on $p_\perp$, 
thanks to a cancellation between   $p_\perp$-dependent contributions to 
individual  helicity amplitudes Eq.(\ref{eqprev}),   when 
the differential cross section  is evaluated \cite{Banfi:2013eda}. 
In principle, it could have been possible to interpret this result as an indication 
that the naive factorization of soft emissions extends to a region beyond $p_\perp > m_b$, 
at least inasmuch as the interference with the top quark loop is concerned. 
However, our result Eq.(\ref{eq:higgspt}) shows that such an interpretation does not hold 
and that the cancellation of $p_\perp$-dependent double logarithmic corrections 
does not persist beyond one-loop. In fact, starting from three loops, the double logarithmic 
corrections to the differential cross section 
start to depend on the rapidity of the emitted gluon  as
Eq.(\ref{eq::fseries}) shows.

To understand numerical impact of these corrections, we use  $m_H =
125~{\rm GeV}$, $m_b = 4.2~{\rm GeV}$, $\alpha_s = 0.12$ and consider $p_\perp$
in the range $ m_b < p_\perp < 50~{\rm GeV}$. We note that the one-loop
double logarithmic corrections reduce the cross section by about $16\%$. This is
somewhat larger than  the result of the full
computation,  but still in the right  ballpark.
The two-loop correction increases the result by about $1.5\%$. This
is somewhat smaller than the next-to-leading order effect in $gg \to H$ cross
section but, given the fact that we only consider the abelian contribution here,
the two results are not inconsistent.\footnote{ 
The  top-bottom interference changes the $m_t \to \infty$ inclusive 
cross section  by approximately 
$-12 \%$ at leading order. QCD corrections to the bottom loop 
decrease this  leading order result by fifty percent.
}  
However, our main interest is in
$p_\perp$-dependent corrections and these corrections turn out to be quite small. In fact,
the two-loop correction in Eq.(\ref{eq:higgspt})  decreases  by just about $0.2
\%$ when the transverse momentum  varies    from  $p_\perp \sim m_b$ to
$p_\perp \sim 50~{\rm GeV}$. This tiny change is the result of a strong
cancellation between $\tau^3$ and $\tau^4$ term in Eq.(\ref{eq:higgspt}).  When
taken separately, these terms could have caused a change in the two-loop result
that is closer to one percent. The three-loop correction in
Eq.(\ref{eq:higgspt}) changes the prediction by about $-0.1\%$ and its
$p_\perp$-dependent part is one order of magnitude smaller.

\section{Conclusion}
\label{sec::conc}
In this paper, we have studied the  bottom-quark loop contribution to the
production of the Higgs boson  in association with a jet in  gluon fusion in the
double logarithmic approximation.  This contribution  is suppressed by  the
ratio of the bottom-quark mass to the Higgs boson mass but, at the same time,  it is
enhanced by two powers of large  logarithms,  $\ln(s/m_b^2)$ or
$\ln(p_\perp^2/m_b^2)$,  per one power of the strong coupling constant. As it  is
repeatedly emphasized in the literature,  these terms may  be
important for phenomenoly, in particular  for the description of the Higgs
boson transverse momentum distribution in an interesting kinematic region $m_b
< p_\perp < m_H$. We have analyzed the  abelian part of the double logarithmic 
corrections and computed  the $gg\to Hg$ helicity
amplitudes which incorporate these terms to all orders in $\alpha_s$.

Numerically, the abelian corrections appear to be moderate. For example, the
two-loop corrections change the transverse momentum distribution  by about two
percent. However it is important to note  that the $p_\perp$-dependent part of
these  corrections is only  about $0.2 \%$ due to the cancellation between
different  $p_\perp$-dependent terms. Assuming that, up to an obvious 
change in the color factor 
$C_F\to
C_A$, the non-abelian corrections 
will be similar to the abelian ones, 
we estimate the yet unknown non-abelian corrections  to be about  three
times larger. We conclude that  the description of  the Higgs boson transverse momentum
distribution with a few percent precision requires a calculation of the ${\cal
O}(\alpha_s)$ logarithmically enhanced non-abelian corrections  to bottom quark contribution 
while the all-order resummation is, probably, not  important. Our analysis sets up a 
framework for such a calculation. A new element  in the calculation of the 
non-abelian part  is its infra-red sensitivity 
and a related need to account for  the  contribution of soft radiation.

\section*{Acknowledgments}
K.M. would like to thank F.~Caola for useful conversations.  The work of A.P. is
supported in part by NSERC and Perimeter Institute of Theoretical Physics.

\appendix

\section{Polarization vectors}
\label{app_pol}

The initial state gluon with momentum $k$ and the gauge vector $r$ is described by 
the following polarization vectors
\be
\epsilon_+^\mu = -\frac{[r \gamma^\mu k \rangle }{\sqrt{2} [r k] },
\;\;\;\;
\epsilon_-^\mu = \frac{ \langle r \gamma^\mu k] }{\sqrt{2} \langle rk \rangle}.
\ee
The polarization vectors for  the final state gluon 
are  obtained by exchanging  $\epsilon_+ \leftrightarrow \epsilon_-$.

As reference vectors, we choose  $p_{1,2}$ for $\epsilon_{2,1}$ and
$p_2$ for $\epsilon_3$. The latter choice allows us to ignore all the
contributions where the soft gluon $g_3$ is emitted by either  gluon $g_2$ or a fermion 
that carries momentum $p_2$.  The full list of polarization vectors that we use in the 
calculation, with all the reference vectors  explicitly shown, reads
\be
\begin{split}
& \epsilon_{+}^{\mu}(1) = -\frac{1}{\sqrt{2}} \frac{[ 2 \gamma^\mu 1 \rangle }{ [2 1] },
\;\;\;\;\;\;
\epsilon_{-}^{\mu}(1) = \frac{1}{\sqrt{2}} \frac{ \langle  2 \gamma^\mu 1 ] }{ \langle 2 1 \rangle  },
\\
& \epsilon_{+}^{\mu}(2) = -\frac{1}{\sqrt{2}} \frac{[ 1 \gamma^\mu 2 \rangle }{ [1 2] },
\;\;\;\;\;\;
\epsilon_{-}^{\mu}(2) = \frac{1}{\sqrt{2}} \frac{ \langle  1 \gamma^\mu 2 ] }{ \langle 1 2 \rangle  },
\\
& \epsilon_{+}^{\mu}(3) = \;\;\frac{1}{\sqrt{2}} \frac{ \langle  2 \gamma^\mu 3 ] }{ \langle 2 3 \rangle  },
\;\;\;\;\;\;
\epsilon_{-}^{\mu}(3) = -\frac{1}{\sqrt{2}} \frac{[ 2 \gamma^\mu 3 \rangle }{ [2 3] }.
\end{split}
\ee


\end{document}